\documentclass[twocolumn,epjc3]{svjour3}          

\def\be{\begin{equation}}
\def\ee{\end{equation}}
\def\bea{\begin{eqnarray}}
\def\eea{\end{eqnarray}}

\usepackage{amssymb}
\usepackage{latexsym}
\RequirePackage{flushend}
\RequirePackage[numbers,sort&compress]{natbib}

\journalname{Eur. Phys. J. C}
\begin{document}



\title{Novel symmetries in Weyl-invariant gravity with massive gauge field}

\author{K. Abhinav\thanksref{e1,addr1} \and A. Shukla\thanksref{e2,addr2}
        \and
        P. K. Panigrahi\thanksref{e3,addr2} 
}

\thankstext{e1}{e-mail: kumar.abhinav@bose.res.in}
\thankstext{e2}{e-mail: as3756@iiserkol.ac.in}
\thankstext{e3}{e-mail: pprasanta@iiserkol.ac.in}

\institute{S.N. Bose National Centre for Basic Sciences, JD Block, Sector III, Salt Lake, Kolkata-700106, India \label{addr1}
           \and
Indian Institute of Science Education and Research Kolkata, Mohanpur-741246, India\label{addr2}
           }

\date{Received: date / Accepted: date}


\maketitle

\begin{abstract}
The background field method is used to linearize the Weyl invariant scalar-tensor
gravity,  coupled with a St\"uckelberg field. For a generic background metric, this action
is found to be {\it not} invariant, under both diffeomorphism and generalized Weyl symmetry,
the latter being a combination of gauge and Weyl transformations. Interestingly, the
quadratic Lagrangian, emerging from a background of Minkowski metric, respects both the
transformations, {\it independently}. Becchi-Rouet-Stora-Tyutin (BRST) symmetry of
scalar-tensor gravity coupled with a St\"uckelberg-like massive gauge particle, possessing 
diffeomorphism and generalized Weyl symmetry, reveals that in both the cases, negative
norm states with unphysical degrees of freedom {\it do} exist. We then show that, by
combining diffeomorphism and generalized Weyl symmetries, all the ghost states decouple,
thereby removing the unphysical redundancies of the theory. During this process, the scalar
field does not represent any dynamic mode, yet modifies the {\it usual} harmonic gauge
condition through non-minimal coupling with gravity.
\end{abstract}





\section{Introduction}
Field theoretic models with scale-invariance have been found useful in the context of various cosmological
problems \cite{Kal,Bars}. Scalar-tensor gravity (STG) is one such theory, which possesses  local
scale-invariance and has been widely used in explaining the inflationary properties of the universe
\cite{kao,zhang,cks}. It has recently been shown that, the conserved current
corresponding to the Weyl symmetry of STG, vanishes identically \cite {Jackiw}. Thus, although the STG
descends from a more general theory of the standard model coupled to gravity \cite{Bars}, at the
level of the corresponding Lagrangian, the pure Weyl symmetry has no dynamical 
role and can be eliminated by making suitable choice of field variables \cite{CCJ,deser}. The {\it physicality}
of the relevant scalar field can be retrieved by considering the {\it geodesic incompleteness}
of such theories \cite{Bars2}. We, however, limit to the STG Lagrangian with no physical mode
associated with the Weyl symmetry. In a recent paper, we have shown that 
there exists a massive scale invariant theory, with non-vanishing generalized scale current, when a St\"uckelberg field is suitably 
coupled with STG. It is imbibed  with a generalized Weyl symmetry: a combination of gauge and local scale symmetry \cite{Own}. 
Remarkably, the STG-St\"ukelberg theory, with gravity non-minimally coupled to the scalar field in STG, is a consistent theory only when the Weyl 
and gauge transformations are inter-related, making STG a part of the bigger theory. 
This aspect may find physical  implication in scale-invariant cosmology.  As is well known,
gravitational theories are notorious to quantize, lacking gauge-fixing with respect to the diffeomorphism redundancy. Therefore, the
classical analysis first needs to be corroborated with a proper gauge-fixing, for
establishing the consistency of the still-extended STG-St\"uckelberg theory.

Quantization of gravity has been a challenge due to the highly non-linear nature 
of gravity. In this process, several important methods e.g., tetrad formalism, background field method etc., have been developed \cite{NO,Carroll}.  
Despite of its limitations due to the inherent simplicity, the background field approach
has widely been exploited in various gravitational models \cite{'tHV,Pagani,Faria}. The 
fact that, the background fields obey classical equations of motion is utilized to attain
the lowest order, tree-level quantum action to be quadratic in nature, yielding equations
of motion, linear in field variables. For the cases of pure gravity \cite{Carroll}, as well
as gravity, minimally coupled to a dilaton scalar field variables \cite{'tHV}, such equations
were obtained by 
identifying proper gauge-fixing against the inherent diffeomorphism redundancy of such systems 
under generic infinitesimal coordinate transformations. Recently, the {\it canonical}
quantization of massive conformal gravity has been performed in the weak quantum
field limit, revealing that the quantum massive gravity is unitary and renormalizable,
but having ghost states \cite{Faria}. Furthermore, using the same background field expansion, a generic Weyl
invariant gravitational model with dilatations, in the presence of a gauge field, has also 
been quantized through suitable definition of renormalization group flows \cite{Pagani}. The 
latter case, including the STG suitably coupled with Abelian gauge field, is of importance in understanding 
light-matter interaction during conformal phases of the universe (e.g., inflation)
and in higher dimensional gravity models with compactification (e.g., Kaluza-Klein
gravity and several string theory models). However, even in these background field models,
exact gauge-fixing of the conformal dilation gravity, coupled to gauge field, 
has not been {\it derived} so far. Further, consistent separation of the unphysical states,
by introduction of proper ghost terms, remains undone till now.

In the present work, we consistently gauge-fix the STG-St\"uckelberg theory \cite{Own},
depicting conformal gravity coupled to massive gauge particle, with the St\"uckelberg scalar field identified 
with the scaling scalar of STG, in the background field method. The
equations of motion, obtained from a quadratic tree-level `quantum' Lagrangian, are linear in 
field variables. Suitable choice of gauge-fixing yields corresponding dynamics with positive 
semi-definite norm. We adopt the celebrated Becchi-Rouet-Stora-Tyutin (BRST) method, where
unitarity and gauge-invariance co-exist \cite{BRS1, BRS2, BRS3,TY1}, to separate-out negative-norm
states, leaving the rest of the dynamic fields `physical'. This is the first instance of 
BRST gauge fixing of such a theory to the best of our knowledge. Though this method allows
for construction of quantum brackets at the tree (Lagrangian) level, it is generally accepted \cite{NO} 
that this further ensures consistent quantization up to all
orders of perturbation. However, presently we leave the proof of the same for later works,
and limit ourselves to the tree-level dynamics, with physical states identified.
To this end, the corresponding Faddeev-Popov \cite{FP} ghosts are
identified, with invariance under {\it two} independent supersymmetric transformations, BRST
and anti-BRST \cite{LB, RPM1, RPM2, RSAM}, which are nilpotent and anticommuting. This enables 
us to identify the {\it effective} physical degrees of freedom of the theory, represented by 
graviton and massive gauge particle, with no dynamics for the scalar field. It is observed that this can be consistent
only when Weyl and gauge symmetries are inter-related with diffeomorphism in a Minkowski background.

The paper is organized as follows. In Sec. 2, we apply the weak field method, with a generic metric as a background 
and linearize the generalized Weyl invariant STG-St\"uckelberg Lagrangian. Further, we show that quadratic Lagrangian 
is neither invariant under the usual diffeomorphism nor under the generalized Weyl transformation. Remarkably, on taking the 
background metric to be Minkowskian, the quadratic Lagrangian regains both the symmetries, which is depicted in Sec. 3. 
Subsequently, Sec. 4 is devoted to the discussion of (anti-)BRST symmetries of the quadratic Lagrangian, emerging from the 
STG-St\"uckelberg theory. We show that, for the removal of 
negative norm states, one has to combine both diffeormophism and generalized Weyl transformations, in an appropriate manner.
The harmonic gauge-fixing condition gets modified to a specific form of the well-known de
Donder gauge due to the non-minimal coupling of gravity with the scalar field.
We conclude after summarizing the obtained result and point out future directions of work.

\section{Background field expansion}
High non-linearity of Einstein's gravity, prevents dynamical
quantization of the same. A full non-perturbative approach has given rise to the $SL(2,\mathbb{R})$
gravity \cite{NO}, described through the tetrad formalism \cite{Ashtekar}. However,
high mathematical complexity has prevented the same from yielding practical 
solutions. Perturbative approaches are relatively simpler, which are based on the physical
assumption of smallness of quantum fluctuations against classical curvature. This is very
much plausible in a macroscopic universe, such as the present one. The background field
approach \cite{'tHV} is applied to expand the complete metric in powers of quantum fluctuations
$h_{\mu\nu}$ \cite{Carroll}, with the corresponding dynamics governed by their second order contributions,
as the first order terms vanish since the classical background fields are always taken to
be on-shell \cite{'tHV}. As the corresponding Lagrangian is second order, the equations of
motion are linear in $h_{\mu\nu}$, leading to linearized gravity \cite{Carroll}.

Quantization of gravity requires, in general, the removal of the inherent
redundancy arising from the symmetry under diffeomorphism transformation,
\bea
\delta_\xi A^{\alpha\beta\cdots}_{\mu\nu\cdots}&=&\xi^\rho\partial_\rho A^{\alpha\beta\cdots}_{\mu\nu\cdots}+\xi^\alpha\partial_\rho A^{\rho\beta\cdots}_{\mu\nu\cdots}+\xi^\beta\partial_\rho A^{\alpha\rho\cdots}_{\mu\nu\cdots}+\cdots \nonumber\\
&-&\xi^\rho\partial_\mu A^{\alpha\beta\cdots}_{\rho\nu\cdots} -\xi^\rho\partial_\nu A^{\alpha\beta\cdots}_{\mu\rho\cdots}.\label{01}
\eea
This makes gravity a gauge theory, with a Fock space containing negative norm (ghost) states,
needing proper gauge-fixing. It is a difficult task in this highly non-perturbative
theory \cite{NO}. In linearized gravity, the obvious approximation of the $SL(2,\mathbb{R})$ 
diffeomorphism transformation does reduces to one, that respects the symmetry of the second
order Lagrangian; but only in the case of pure gravity \cite{Carroll}. In general, the
presence of other fields coupled to gravity, that also transform under diffeomorphism, do not
allow the same to be the symmetry of the full theory. This makes the theory non-covariant
over general coordinate transformation. Therefore, very restricted boundary conditions are
required \cite{Faria, Pagani} to quantize a linear gravitational model with generic matter coupling, 
while preserving covariance over diffeomorphism.

As a primary need for quantization, by separating physical degrees of freedom 
from the unphysical ones, We proceed to gauge-fix the combined STG-St\"uckelberg theory \cite{Own},
\bea
{\cal L}_{\rm STSt}&=&\sqrt{-{\cal G}}\Big[\frac{1}{12\kappa}{\cal R}\Phi^2+\frac{1}{2\kappa}\partial_\mu\Phi\partial_\nu\Phi {\cal G}^{\mu\nu}\nonumber\\
&-&\frac{1}{4}{\cal F}_{\mu\alpha}{\cal F}_{\nu\beta}{\cal G}^{\mu\nu}{\cal G}^{\alpha\beta}+\frac{m^2}{2}{\cal G}^{\mu\nu}{\cal D}_\mu\Phi{\cal D}_\nu\Phi\Big],\label{JPSt2}
\eea
depicting a generalized Weyl-invariant massive gauge field in presence of curvature. Here, $\kappa = (16\pi G)$
is a constant of dimension inverse of square of the mass, with $G$ being Newtonian gravitational 
constant, and $R$ is the Ricci scalar. Such a model can
be useful in explaining the behavior of a conformal, scale invariant universe having short-range
gauge interactions. Here, the scalar field $\Phi$ couples non-minimally to gravity, defined by
the metric ${\cal G}_{\mu\nu}$, in the first term of the above Lagrangian. As explained above,
quantization of the above theory is much more difficult than Einstein-Hilbert gravity, given the 
nature of different couplings, including that of the Abelian gauge field $A_\mu$ represented by
the corresponding field tensor ${\cal F}_{\mu\nu}$. Further, it should be noted that ${\cal D}_\mu=\partial_\mu-A_\mu$,
a non-$U(1)$ covariant derivative corresponding to the gauge connection \cite{Own}. We will opt
for quadratic form of this action, following the background field expansions:
\bea
{\cal G}_{\mu\nu}&=&g_{\mu\nu}+h_{\mu\nu}=g_{\mu\rho}\left(\delta^\rho_\nu+h^\rho_\nu\right),\nonumber\\
{\cal G}^{\mu\nu}&=&g^{\mu\nu}-h^{\mu\nu}+h^\mu_\rho h^{\rho\nu},\nonumber\\
h^{\mu\nu}&=&h^{\nu\mu},\qquad{\cal G}_{\mu\nu}{\cal G}^{\nu\rho}=\delta^\rho_\mu=g_{\mu\nu}g^{\nu\rho};\nonumber\\
A_\mu&=&a_\mu+b_\mu;\qquad\Phi=\varphi+\phi.\label{1}
\eea
The first terms in the above are classical fields, obeying classical equations of motion (EOM)
\cite{Own},
\bea
\nabla_\mu F^{\mu\nu}&=&-m^2\varphi^2a^\nu+m^2\varphi\partial^\nu\varphi,\nonumber\\
\Box\varphi&=&\frac{1}{1+\kappa m^2}\left[\frac{1}{6}R+\kappa m^2\left(\nabla_\mu a_\nu+a_\mu a_\nu\right)g^{\mu\nu}\right]\varphi,\nonumber\\
{\cal T}_{\mu\nu}&=&\frac{1}{\kappa}\Bigl[\frac{1}{12}G_{\mu\nu}\varphi^2+\frac{1}{2}\partial_\mu\varphi\partial_\nu\varphi-\frac{1}{4}g_{\mu\nu}g^{\alpha\beta}\partial_\alpha\varphi\partial_\beta\varphi \nonumber\\
&+&\frac{1}{12}\left(g_{\mu\nu}\Box-\nabla_\mu\nabla_\nu\right)\varphi^2\Bigr]
-\frac{1}{2}F_\mu^{~\alpha}F_{\nu\alpha} \nonumber\\ 
&+&\frac{m^2}{2}\varphi^2\bar{a}_\mu\bar{a}_\nu 
-\frac{1}{4}g_{\mu\nu}\Bigl(-\frac{1}{2}F_{\alpha\beta}F^{\alpha\beta}\nonumber\\
&+&m^2\varphi^2\bar{a}_\alpha\bar{a}^\alpha\Bigr)=0;\label{EOM02}
\eea
with,
\bea
&&\Box\varphi=\frac{1}{\sqrt{-g}}\partial_\mu\left[\sqrt{-g}g^{\mu\nu}\partial_\nu\varphi\right],\quad \bar{a}_\mu=a_\mu-\partial_\mu\log\varphi,\nonumber\\
&& G_{\mu\nu}=R_{\mu\nu}-\frac{1}{2}g_{\mu\nu}R.\label{Notation1}
\eea
The second terms in Eqs. (\ref{1}) are quantum fluctuations, which are orders of magnitude 
suppressed. Here, we will follow the treatment by 't Hooft and Veltman \cite{'tHV}, carried out
for gravity minimally coupled with a scalar field, to obtain the most general second order quantum
contribution to the tree-level Lagrangian of Eq. (\ref{JPSt2}).

The most non-trivial term of the action, in view of background field expansion, is
the first one, with gravity non-minimally coupled to the scalar field, which will be treated in
detail. The background (classical) metric $g_{\mu\nu}$ raises and lowers indices and
non-trivially contributes to the background covariant derivative $D_\mu$. The gravitational
Jacobian expands as \cite{'tHV},
\bea
\sqrt{-{\cal G}}&=&\sqrt{-g}\left(1+\frac{1}{2}h-\frac{1}{4}h^{\mu\nu}h_{\mu\nu}+\frac{1}{8}h^2\right); \nonumber\\ 
h &=& h^\mu_\mu,\label{2}
\eea
where we have used the second expression from the first equation of Eqs. (\ref{1}). Then the first 
term in Eq. (\ref{JPSt2}) can be expanded, in the powers of the `quantum' metric $h_{\mu\nu}$ and 
scalar field $\phi$, in the form,
\be
{\cal L}_g:=\frac{1}{12\kappa}\sqrt{-{\cal G}}{\cal R}\Phi^2\equiv{\cal L}_0+{\cal L}_1+{\cal L}_2+{\cal O}(3),\label{11}
\ee
wherein ${\cal L}_0$ has no `quantum dynamics'. ${\cal L}_1$ is first order in $h_{\mu\nu}$ and
$\phi$, and its contribution is exactly canceled by that coming from the first order contribution
from the rest of the Lagrangian, following the classical equations of motion \cite{'tHV}. Hence,
the lowest order quantum dynamics arises from ${\cal L}_2$, which is quadratic in quantum
fields, and can be decomposed into the three parts ${\cal L}_2={\cal L}_a+{\cal L}_b+{\cal L}_c$,
with the expressions,
\bea
&&{\cal L}_a=\frac{\sqrt{-g}}{12\kappa}{^0R}\phi^2,\nonumber\\
&&{\cal L}_b=\frac{\sqrt{-g}}{6\kappa}\left(h^{\mu\nu}_{~~;\mu\nu}-\Box h+h^{\mu\nu~}{^0R_{\mu\nu}}-\frac{1}{2}h{^0R}\right)\varphi\phi,\nonumber\\
&&{\cal L}_c=\frac{\sqrt{-g}}{12\kappa}\Big[\frac{1}{8}\left(h^2-2h^{\mu\nu}h_{\mu\nu}\right){^0R}-h^{\alpha\mu}h_{\mu\nu}{^{~0}R^\nu_\alpha}\nonumber\\
&&+\frac{1}{2}hh^{\mu\nu}{^{~0}R_{\mu\nu}}+\frac{1}{4}h^{\mu\nu}\Box h_{\mu\nu}-\frac{1}{4}h\Box h +\frac{1}{2}hh^{\mu\nu}_{~~;\mu\nu}\nonumber\\
&&+\frac{1}{2}h^{\mu\nu;\alpha}h_{\mu\alpha;\nu}\Big]\varphi^2.\label{12}
\eea
The last expression is obtained through integration by parts and removing total derivatives, 
with commutator for covariant derivatives being taken into account. Here, the semi-colon indicates
covariant derivative with respect to the background metric. Moreover,  ${^{~0}R_{\mu\nu}} $ and ${^0R}$ represent the Ricci tensor 
and scalar with respect to the background metric, respectively.

Expanding the rest of the Lagrangian in Eq. (\ref{JPSt2}), we get,
\bea
{\cal L}^{(2)}_{\rm STSt}&\equiv&{\cal L}_2+\frac{\sqrt{-g}}{2\kappa}\Big[\frac{1}{8}\left(h^2
-2h^{\mu\nu}h_{\mu\nu}\right)\varphi_{,\alpha}\varphi^{,\alpha} \nonumber\\
&+&\frac{h}{2}\big(2\varphi_{,\mu}\phi^{,\mu} 
-h^{\mu\nu}\varphi_{,\mu}\varphi_{,\nu}\big)+\varphi_{,\mu}h^\mu_{~\alpha}h^{\alpha\nu}\varphi_{,\nu} \nonumber\\
&-&2h^{\mu\nu}\varphi_{,\mu}\phi_{,\nu}+\phi_{,\mu}\phi^{,\mu}\Big] -\frac{\sqrt{-g}}{4}\Big[\frac{1}{8}\big(h^2 \nonumber\\ 
&-&2h^{\mu\nu}h_{\mu\nu}\big)A^{\alpha\beta}A_{\alpha\beta} + h\Big(A_{\mu\nu}B^{\mu\nu} \nonumber\\
&-& h^{\mu\nu}A_\mu^{~\alpha}A_{\nu\alpha}\Big)  
+ 2A_{\mu\alpha}h^\mu_{~\beta}h^{\beta\nu}A^\alpha_{~\nu} \nonumber\\
&+& h^{\mu\nu}h^{\alpha\beta}A_{\mu\alpha}A_{\nu\beta} + B_{\mu\nu}B^{\mu\nu} -2h^{\mu\nu}A_\mu^{~\alpha}B_{\nu\alpha}\Big] \nonumber\\
&+& \frac{\sqrt{-g}}{2}m^2\Big[\frac{1}{8}\big(h^2 
-2h^{\mu\nu}h_{\mu\nu}\big)d_\alpha\varphi d^\alpha\varphi + d_\mu\phi d^\mu\phi \nonumber\\
&-&\frac{1}{2}hh^{\mu\nu}d_\mu\varphi d_\nu\varphi  
+ d_\mu\varphi h^\mu_{~\alpha}h^{\alpha\nu}d_\nu\varphi 
-2b_\mu\varphi d^\mu\phi \nonumber\\
&-& b_\mu\phi d^\mu\varphi +\left(h g^{\mu\nu} - 2h^{\mu\nu}\right)\left(d_\mu\varphi d_\nu\phi
-b_\mu\varphi d_\nu\varphi\right) \nonumber\\
&+&b_\mu b^\mu\varphi^2 \Big] + {\cal O}(h^3) \nonumber\\
&=&{\cal L}_2+{\cal L}_\phi+{\cal L}_{\rm gauge}+{\cal L}_{\rm int}.\label{QL}
\eea
This action is quadratic in quantum fields and represents their dynamics. Here, $A_{\mu\nu}$
and $B_{\mu\nu}$ correspond to the background and quantum gauge field strengths, respectively and
$d_\mu:=\partial_\mu-a_\mu$.

The present theory, being coupled to gravity, is expected to be invariant under 
infinitesimal general coordinate transformation of the form,
\be
x^\mu\rightarrow x^\mu+\xi^\mu,\label{N04}
\ee
with $\xi^\mu$ being small. The corresponding variation of the fields is diffeomorphism, under which,
background fields do not change, whereas the quantum fields do \cite{'tHV,Carroll,Faria} as
\cite{'tHV,Carroll},
\bea
h_{\mu\nu}&\rightarrow& h_{\mu\nu}+\xi_{(\mu;\nu)}+\xi^\alpha_{~;(\mu}h_{\nu)\alpha}+\xi^\alpha h_{\mu\nu;\alpha}, \nonumber\\
 h^{\mu\nu}&\rightarrow & h^{\mu\nu}-\xi^{(\mu;\nu)}-\xi^{\alpha;(\mu}h^{\nu)}_\alpha-\xi^\alpha h^{\mu\nu}_{~~;\alpha},\nonumber\\
h&\rightarrow& h+2\xi^\mu_{~;\mu}+2h^{\mu\nu}\xi_{\mu;\nu}+\xi^\mu h_{;\mu},\nonumber\\
b_\mu&\rightarrow&b_\mu+\xi^\rho\partial_\rho A_\mu+A_\rho\partial_\mu\xi^\rho,\nonumber\\
\phi &\rightarrow & \phi+\xi^\rho\partial_\rho\Phi.\label{dif01}
\eea
Here $\xi^\mu$ is identified as the diffeomorphism parameter. As observed earlier \cite{'tHV}, if smallness
of the quantum fields are {\it not} taken into account, {\it i.e.} the Lagrangian being 
considered up to all orders of expansion, the transformations in Eqs. (\ref{dif01})
represent symmetry of the full Lagrangian, not only that of the quadratic part. Since
the present goal is concerned with the second order contribution, yielding linear EOMs, one needs
to reduce these transformations suitably up to certain orders. The obvious choice is to restrict
the R.H.S. of Eqs. (\ref{dif01}) to zero order in quantum fields, as $\xi_\mu$ is small
\cite{Carroll}. However, unlike pure gravity, the presence of other interacting dynamic fields
in the present case forbids such a reduced symmetry of the second order Lagrangian in Eq. (\ref{QL}).

The same is true for the independent generalized Weyl transformations \cite{Own},
which can be attributed completely to the quantum fields \cite{Faria} as,
\be
h_{\mu\nu}\rightarrow h_{\mu\nu}-2\theta{\cal G}_{\mu\nu},\quad b_\mu\rightarrow b_\mu+\partial_\mu\theta,\quad\phi\rightarrow\phi+\theta\Phi,\label{GW01}
\ee
which is a symmetry of the full Lagrangian in Eq. (\ref{JPSt2}). For the second order Lagrangian,
this is no more a symmetry, even when approximated to zero order in quantum fields. This imposes
strict restrictions on the background fields, in order to retain both diffeomorphism and generalized Weyl symmetries, 
in reduced forms, in the Lagrangian. In the following, we obtain such
a second order Lagrangian with specific background (classical) fields.

\section{Second order theory with constant background}
Following the case of pure gravity \cite{Carroll} and that of conformal scalar-tensor gravity
\cite{Faria}, wherein a reduced diffeomorphism is obtainable in the second order theory for
a constant (Minkowski) background metric, for the present case, we consider all three background
fields to be constant to yield,
\bea
{\cal G}_{\mu\nu} = \eta_{\mu\nu} + h_{\mu\nu}, \qquad \Phi = \varphi + \phi, \qquad a_\mu  = b_\mu, \nonumber\\
\sqrt -g = 1 + \frac{1}{2} h - \frac{1}{4} h^{\mu\nu} h_{\mu\nu} + \frac{1}{8} h^2 + ...,
\eea
where $\eta_{\mu\nu}$ is the Minkowski metric and $a_\mu=0$. The latter, more restrictive choice owes to the fact that a constant
non-zero, non-dynamic vector field destroys the isotropy of the background space, when coupled to the dynamic fields.
Further, the generalized Weyl symmetry is broken in presence of a vector that does not transform. These approximations
simplify the second order Lagrangian, further as,
\bea
{\cal L}^{(2){\rm red}}_{\rm STSt} &=& \frac{1}{12\kappa}\Big[\frac{1}{4}h_{\mu\nu}\Box h^{\mu\nu}+\frac{1}{2}h \partial_\mu\partial_\nu h^{\mu\nu}
+\frac{1}{2} \partial_\alpha h^{\mu\alpha} \partial^\beta h_{\mu\beta} \nonumber\\
&-&\frac{1}{4}h\Box h\Big]\varphi^2
+\frac{1}{6\kappa}\Big[\partial_\mu\partial_\nu h^{\mu\nu} - \Box h\Big]\varphi\phi \nonumber\\ 
&+& \frac{1}{2\kappa}\partial_\mu \phi \partial^\mu \phi
-\frac{1}{4}B_{\mu\nu}B^{\mu\nu}+\frac{1}{2}m^2\Big[\partial_\mu\phi\partial^\mu\phi\nonumber\\ &-&2b_\mu\varphi\partial^\mu\phi 
+ b_\mu b^\mu\varphi^2\Big].\label{CBQL2}
\eea
Under the reduced diffeomorphism transformations,
\be
\delta_\xi h_{\mu\nu} = \partial_\mu \xi_\nu + \partial_\nu \xi_\mu ,\qquad \delta_\xi b_\mu = 0, \qquad  \delta_\xi\phi =0,\label{23}
\ee
with quantum fields $b_\mu$ and $\phi$ unchanged, the simplified second order Lagrangian is
invariant, as it changes by a total derivative term,
\bea
&\delta_\xi {\cal L}^{(2){\rm red}}_{\rm STSt} = \partial_\mu \Bigl[ \frac{1}{24\kappa} \Big\{(\partial_\nu \xi_\alpha) \partial^\mu h^{\nu\alpha} 
-  \partial^\mu (\partial_\nu \xi_\alpha) h^{\nu\alpha} \nonumber\\
&-(\partial_\alpha \xi^\alpha) \partial^\mu h + \partial^\mu(\partial_\alpha \xi^\alpha) h \nonumber\\ 
&+ 2 (\partial_\alpha \xi^\alpha) \partial_\nu h^{\mu\nu} \Big\} \varphi^2\Bigr].
\eea
The generalized Weyl transformation for the quantum fields reduces to, 
\be
\delta_\theta h_{\mu\nu} = -2\theta\eta_{\mu\nu},\qquad\delta_\theta b_\mu = \partial_\mu \theta,\qquad\delta_\theta\phi = \theta\varphi,\label{24}
\ee
which also leaves the Lagrangian in Eq. (\ref{CBQL2}) invariant modulo a total derivative,
\bea
\delta_\theta {\cal L}^{(2){\rm red}}_{\rm STSt} &=&  \partial_\mu\Bigl[ \frac{1}{24\kappa}  \bigl ( h \partial^\mu \theta 
- \partial^\mu h \theta - 4 \theta \partial_\nu h^{\mu\nu}\bigr)  \varphi^2 \nonumber\\ 
&+& \frac{1}{\kappa}\varphi \phi \partial^\mu \theta \Bigr].
\eea
As both diffeomorphism and generalized Weyl transformations are independent, they can be combined
as,  
\bea
&&\delta h_{\mu\nu} = \partial_\mu \xi_\nu + \partial_\nu \xi_\mu - 2 \eta_{\mu\nu} \theta, \nonumber\\ 
&&\delta b_\mu = \partial_\mu\theta, 
\qquad \delta \phi = \theta \varphi,\label{25}
\eea
with $\delta = \left(\delta_\xi+\delta_\theta\right)$, leaving the corresponding Lagrangian
invariant modulo a total derivative term,
\bea
\delta {\cal L}^{(2){\rm red}}_{\rm STSt} &=&  \partial_\mu\Bigl[ \frac{1}{24\kappa} \Big\{(\partial_\nu \xi_\alpha) \partial^\mu h^{\nu\alpha} 
-  \partial^\mu (\partial_\nu \xi_\alpha) h^{\nu\alpha} \nonumber\\
&-& (\partial_\alpha \xi^\alpha) \partial^\mu h
 +  \partial^\mu(\partial_\alpha \xi^\alpha) h + 2 (\partial_\alpha \xi^\alpha) \partial_\nu h^{\mu\nu} \Big\}\varphi^2 \nonumber\\
&+& \frac{1}{24\kappa}  \bigl ( h \partial^\mu \theta 
- \partial^\mu h \theta - 4 \theta \partial_\nu h^{\mu\nu}\bigr)  \varphi^2 \nonumber\\
&+& \frac{1}{\kappa}\varphi \phi \partial^\mu \theta \Bigr].
\eea
As the Lagrangian in (\ref{CBQL2}) is invariant, independently under approximated 
diffeomorphism and generalized Weyl transformations, {\it both} of which represent constraints,
the physical degrees of freedom of this system is less than the number of field components.
Direct quantization of such a system will lead to a Fock space with negative norm (energy) states,
corresponding to unphysical degrees of freedom. This calls for gauge-fixing each of the two
symmetries, so that the equations of motion corresponding to {\it each} field correspond only to
positive energy states. To this end, the proper gauge-fixing conditions are to be obtained from
the respective Euler-Lagrange equations for fields $h_{\mu\nu}, b_\mu$ and $\phi$, which are,
\bea
&& \Box h^{\mu\nu}+\eta^{\mu\nu} \partial_\alpha \partial_\beta h^{\alpha\beta}- \eta^{\mu\nu}\Box h
+ \partial^\mu \partial^\nu h + \partial_\alpha \partial^\mu h^{\alpha\nu} \nonumber\\ &&+ \partial_\alpha \partial^\nu h^{\alpha\mu} 
=\frac{4}{\varphi}\Big(\eta^{\mu\nu}\Box\phi- \partial^\mu\partial^\nu \phi\Big),\nonumber\\
&& \frac{1}{6\kappa}\left( \partial_\mu \partial_\nu h^{\mu\nu} - \Box h\right)\varphi = \left(\frac{1}{\kappa}
+ m^2\right)\Box\phi- m^2\varphi \partial_\mu b^\mu,\nonumber\\
&& \partial_\mu F^{\mu\nu}= m^2\left(\varphi \partial^\nu \phi-\varphi^2 b^\nu\right).\label{QEOM1}
\eea
In the  gravity sector, the {\it extended} harmonic gauge choice \cite{'tHV},
\be
\partial_\mu h^{\mu\nu}-\frac{1}{2}\partial^\nu h - \frac{2}{\varphi}\partial^\nu\phi = 0
,\label{28}
\ee 
leads to the EOM for $h_{\mu\nu}$ as,
\be
\Box \bar{H}^{\mu\nu}=0;\qquad \bar{H}^{\mu\nu}:=h^{\mu\nu}-\frac{1}{2}\eta^{\mu\nu}h-\frac{2}{\varphi}\eta^{\mu\nu}\phi.\label{27}
\ee
Thus, the `gravitational' mode is represented by a linear combination of quantum metric 
and scalar fields, which is expected from the non-minimal coupling between the two \cite{'tHV,CCJ,deser}. 
This physically makes sense, as the scalar mode correspond to both scaling and
St\"uckelberg mechanism \cite{Own}. Hence, it should not represent any physical degree of freedom.
This assertion will be clearer in the following and also in the next section, when (anti-) BRST
symmetry will be obtained.

In the vector gauge sector, the independent covariant gauge choice of,
\be
\partial_\mu b^\mu + m^2\varphi\phi=0,\label{30}
\ee
much like the St\"uckelberg case \cite{Das}, leads to the EOM for $b_\mu$ as,
\be
\left(\Box+m^2\varphi^2\right)b^\nu=0.\label{29}
\ee
This particular gauge choice is considered due to the fact that, the gauge field couples to the scalar one, as in
Eq. \ref{JPSt2}, just like the St\"uckelberg theory, with the coupling terms re-arranged. This 
further justifies acquired mass,
\be
m_b = m\varphi,\label{32a}
\ee
of the gauge field. Further, on using the gauge-fixing conditions in Eqs. (\ref{28}) and (\ref{30}),
and quantum metric EOM of Eq. (\ref{27}) thereby substituting $h_{\mu\nu}$, $h$ and $b_\mu$, the EOM for $\phi$ reduces to,
\be
\left(\Box + m^2\varphi^2\right)\phi = 0,\label{31}
\ee
with a `mass term',
\be
m_\phi = m\varphi\equiv m_b.\label{32b}
\ee
The last result is of importance, as it ensures that Eq. (\ref{31}) can be obtained from Eq.
(\ref{29}) subjected to the gauge-fixing condition of Eq. (\ref{30}). However, the reverse is
{\it not} true. This further shows that $\phi$ does not represent an independent dynamical mode,
as asserted earlier. The two gauge-fixing conditions relate the scalar field independently to the
tensor and vector fields. Also, {\it both} these gauge-fixing conditions are required to obtain
Eq. (\ref{31}), and it can be written purely in terms of other two fields, which is {\it not} the
case for Eqs. (\ref{27}) and (\ref{29}). Another way to validate this point is that the Lagrangian
in Eq. (\ref{JPSt2}) was obtained through a generalized Weyl scaling of a Proca action in presence 
of gravity \cite{Own}, with the scalar component $\Phi$ exactly being the scaling field \cite{Jackiw}
and also the St\"uckelberg-type field. In both regards, it is expected that the scalar component
will not yield any dynamics.

Therefore, we have obtained a massless gravitational and  massive 
vector field at the tree-level, subjected to the gauge-fixing conditions in Eqs. (\ref{28}) and (\ref{30}).
This is in accordance to the earlier observations \cite{'tHV, CCJ, deser},
with modification coming due to the additional gauge coupling \cite{Own}. As symmetric
$h_{\mu\nu}$ has 10 independent components and the $b_\mu$
has 4; 4 modified harmonic constraints Eq. (\ref{28}), 4 coming from the generic coordinate
transformation Eq. (\ref{N04}) \cite{Carroll} and 1 vector gauge constraint Eq. (\ref{30}) leave the net 
physical degrees of freedom of the system to be $(10+4)-(4+4+1)=5$, with 2 for graviton and 3 for
massive gauge particle, which is expected physically. This aspect will be further elucidated, while
introducing the ghost fields for the theory, in the next section. As the exact gauge-fixing conditions have been
achieved, by obtaining `physical' equations of motion (\ref{27}) and (\ref{29}) with positive spectra, 
the (anti-)BRST symmetries of the present system can be carried out, which is demonstrated in the next
section.

\section{(Anti-)BRST transformations} 
Our goal is to identify the physical degrees of freedom of the present theory, which has
redundancies arising from both diffeomorphism and generalized Weyl transformations. To this end, we 
opt for the BRST approach, wherein auxiliary gauge-fixing conditions are 
chosen at the expense of introducing Faddeev-Popov ghost and anti-ghost fields, representing
negative norm states arising from the gauge redundancies of the original theory \cite{FP}. Effectively, this
increase in variables re-locates the over-counting of degrees of freedom from the gauge to the
(anti-)ghost sector, as gauge transformation is replaced by the corresponding (anti-)BRST 
transformations \cite{BRS1,BRS2,BRS3,TY1}. The latter is nilpotent and as it is done off-shell, one is left with only
physical degrees of freedom in the gauge sector, defined in a Fock space with positive
semi-definite norm.

In the present case, the diffeomorphism gauge sector contains only one vector ghost
term in the Lagrangian, since it is confined only to $h_{\mu\nu}$, with $\phi$ being non-dynamic.
Similarly, the generalized Weyl redundancy will lead to {\it one} scalar ghost term, that affects
all the three fields, including $b_\mu$ and $\phi$. This is because the gauge-fixing term
for $h_{\mu\nu}$ remains invariant under the corresponding nilpotent (anti-)BRST transformations. Finally,
the combined diffeomorphism and generalized Weyl gauge sector is dealt with, achieving the physical
states of the full theory, through necessary inclusion of both vector and scalar (anti-)ghost 
fields.

\subsection{For diffeomorphism symmetry}
Here, we consider the path-integral (anti-)BRST symmetry of the Lagrangian with respect to diffeomorphism.
This is exclusive to the gravitational sector, through introduction of corresponding
vector (anti-)ghost fields, made possible by its independence from the generalized Weyl
symmetry in the present theory. For this purpose, we gauge-fix only the gravitational sector of
the complete Lagrangian in Eq. (\ref{CBQL2}), in the modified harmonic gauge of Eq. (\ref{28}) and introduce corresponding 
vector (anti-)ghost fields ($\bar{c}_\mu$)$c_\mu$, with ghost numbers
(-1)1, to yield,  
\bea
{\cal L}^{(2)}_{\rm GF1}&=& \frac{1}{12\kappa}\Big[\frac{1}{4}h_{\mu\nu}\Box h^{\mu\nu}+\frac{1}{2}h \partial_\mu\partial_\nu h^{\mu\nu}
+\frac{1}{2} \partial_\alpha h^{\mu\alpha} \partial^\beta h_{\mu\beta} \nonumber\\
&-&\frac{1}{4}h\Box h\Big]\varphi^2
+\frac{1}{6\kappa}\Big[\partial_\mu\partial_\nu h^{\mu\nu} - \Box h\Big]\varphi\phi +\frac{1}{2\kappa}\partial_\mu \phi \partial^\mu \phi \nonumber\\
&-&\frac{1}{4}B_{\mu\nu}B^{\mu\nu}  +\frac{1}{2}m^2\Big[\partial_\mu\phi\partial^\mu\phi
- 2 b_\mu\varphi\partial^\mu\phi + b_\mu b^\mu\varphi^2\Big] \nonumber\\
&-& \frac{1}{24 \kappa} \Bigl(\partial_\nu h^{\mu\nu}-\frac{1}{2} \partial^\mu h - 2\varphi^{-1}\partial^\mu \phi\Bigr)^2 \nonumber\\
&+& \frac{i}{12\kappa}\partial^\mu {\bar c}^\nu \partial_\mu c_\nu. \label{BRST11}
\eea
Here, and afterwards, we explicitly use the Feynman-'t Hooft gauge. The above Lagrangian can be
linearized by adding the Nakanishi-Lautrup type auxiliary vector field $B_\mu$ as,
\bea
{\cal L}^{(2)}_{\rm GF1}&\equiv& \frac{1}{12\kappa}\Big[\frac{1}{4}h_{\mu\nu}\Box h^{\mu\nu}+\frac{1}{2}h \partial_\mu\partial_\nu h^{\mu\nu}
+\frac{1}{2} \partial_\alpha h^{\mu\alpha} \partial^\beta h_{\mu\beta} \nonumber\\
&-&\frac{1}{4}h\Box h\Big]\varphi^2
+\frac{1}{6\kappa}\Big[\partial_\mu\partial_\nu h^{\mu\nu} - \Box h\Big]\varphi\phi \nonumber\\
&+&\frac{1}{2\kappa}\partial_\mu \phi \partial^\mu \phi
-\frac{1}{4}B_{\mu\nu}B^{\mu\nu}  +\frac{1}{2}m^2\Big[\partial_\mu\phi\partial^\mu\phi \nonumber\\
&-& 2 b_\mu\varphi\partial^\mu\phi 
+ b_\mu b^\mu\varphi^2\Big] + \frac{1}{24 \kappa} B_\mu B^\mu \nonumber\\ 
&+& \frac{1}{12\kappa} B_\mu \Bigl(\partial_\nu h^{\mu\nu}-\frac{1}{2} \partial^\mu h - 2\varphi^{-1}\partial^\mu \phi \Bigr) \nonumber\\
&+& \frac{i}{12\kappa}\partial^\mu {\bar c}^\nu \partial_\mu c_\nu .\label{BRST12}
\eea
with $B_\mu = -\Big(\partial_\nu h^{\mu\nu}-\frac{1}{2} \partial^\mu h - 2\varphi^{-1}\partial^\mu \phi \Big)$
on-shell. The action corresponding to the above Lagrangian is invariant under the following
off-shell nilpotent (anti-)BRST transformations,
\bea
&& s^{(1)}_b h_{\mu\nu} = \partial_\mu c_\nu + \partial_\nu c_\mu, \quad s^{(1)}_b c_\mu = 0,\quad  s^{(1)}_b {\bar c}_\mu = i B_\mu, \nonumber\\
&&  s^{(1)}_b B_\mu = 0 \qquad s^{(1)}_b b_\mu = 0, \qquad  s^{(1)}_b \phi = 0,\nonumber\\
&& s^{(1)}_{ab} h_{\mu\nu} = \partial_\mu {\bar c}_\nu + \partial_\nu {\bar c}_\mu, \quad  s^{(1)}_{ab} {\bar c}_\mu = 0,\quad
 s^{(1)}_{ab} c_\mu = - i B_\mu, \nonumber\\ &&  s^{(1)}_b B_\mu = 0\qquad s^{(1)}_{ab} b_\mu = 0, \qquad s^{(1)}_{ab} \phi = 0, 
\eea 
as the Lagrangian transforms by total derivative terms, respectively, as,
\bea
 s^{(1)}_b {\cal L}^{(2)}_{\rm GF1} &=& \partial_\mu\Bigl [\frac{1}{24\kappa} \Big\{(\partial_\nu c^\alpha) \partial^\mu h^{\nu\alpha} -  \partial^\mu (\partial_\nu c_\alpha) h^{\nu\alpha} \nonumber\\ &-& (\partial_\alpha c^\alpha) \partial^\mu h
+  \partial^\mu(\partial_\alpha c^\alpha) h + 2 (\partial_\alpha c^\alpha) \partial_\nu h^{\mu\nu} \Big\}\varphi^2 \nonumber\\ 
&+& \frac{1}{12\kappa}B^\nu \partial^\mu c_\nu \Bigr], \nonumber\\
 s^{(1)}_{ab} {\cal L}^{(2)}_{\rm GF1} &=& \partial_\mu\Bigl [\frac{1}{24\kappa} \Big\{(\partial_\nu {\bar c}^\alpha) \partial^\mu h^{\nu\alpha} 
-  \partial^\mu (\partial_\nu {\bar c}_\alpha) h^{\nu\alpha} \nonumber\\ 
&-& (\partial_\alpha {\bar c}^\alpha) \partial^\mu h 
+  \partial^\mu(\partial_\alpha {\bar c}^\alpha) h + 2 (\partial_\alpha {\bar c}^\alpha) \partial_\nu h^{\mu\nu} \Big\}\varphi^2 \nonumber\\
&-& \frac{1}{12\kappa}B^\nu \partial^\mu {\bar c}_\nu \Bigr]. 
\eea
Therefore, the gauge redundancy corresponding to diffeomorphism is removed by the inclusion of
the gauge-fixing term, whereas the corresponding symmetry is now compensated by the (anti-)BRST
symmetry of the (anti-)ghost fields. The physical credibility of the above procedure is
reflected by the fact that, in the original Lagrangian of Eq. (\ref{CBQL2}), the symmetric
tensor field $h_{\mu\nu}$ had 10 degrees of freedom, unlike the physical graviton, that
has only 2. To identify the correct number, one has to go to the level of equations of motion
(on-shell) to figure out 8 constraint equations, 4 from harmonic gauge and the other four from the 
remaining choice of the corresponding parameter $\xi_\mu$ in coordinate transformation, as discussed in the previous section. 
On the other hand, introduction of vector (anti-)ghost fields, with four {\it negative}
degrees of freedom each, corresponds to the same physics at the level of the Lagrangian itself
(off-shell). The auxiliary field is always on-shell and does not have any dynamics. This enables
$h_{\mu\nu}$ in ${\cal L}^{(2)}_{\rm GF1}$ to represent exact physical degrees of freedom and
thus, with only positive energy states in the Fock space,
with the negative energy ones attributed to the (anti-)ghost fields.

It is to be noted that, only the gravitational sector is considered above, mainly 
to demonstrate the physical aspects of the same. Eq. (\ref{BRST12}) represents a two-component 
tensor field, with semi-positive norm, coupled to the fields $b_\mu$, with gauge redundancy,
and $\phi$. Thus, only the prior can represent tree-level quantum dynamics. To achieve the full spectrum
of physical states for ${\cal L}^{(2){\rm red}}_{\rm STSt}$, we need to fix the generalized Weyl redundancy
corresponding to the other two fields, which is done later.

\subsection{For generalized Weyl symmetry}
In order to gauge-fix the generalized Weyl symmetry, which includes tensor, vector and scalar sectors
of ${\cal L}^{(2){\rm red}}_{\rm STSt}$, unlike diffeomorphism before, we extend ${\cal L}^{(2){\rm red}}_{\rm STSt}$ to,  
\bea
{\cal L}^{(2)}_{\rm GF2}&=& \frac{1}{12\kappa}\Big[\frac{1}{4}h_{\mu\nu}\Box h^{\mu\nu}+\frac{1}{2}h \partial_\mu\partial_\nu h^{\mu\nu}
+\frac{1}{2} \partial_\alpha h^{\mu\alpha} \partial^\beta h_{\mu\beta} \nonumber\\
&-&\frac{1}{4}h\Box h\Big]\varphi^2
+\frac{1}{6\kappa}\Big[\partial_\mu\partial_\nu h^{\mu\nu} - \Box h\Big]\varphi\phi  
+\frac{1}{2\kappa}\partial_\mu \phi \partial^\mu \phi \nonumber\\
&-&\frac{1}{4}B_{\mu\nu}B^{\mu\nu}  +\frac{1}{2}m^2\Big[\partial_\mu\phi\partial^\mu\phi- 2 b_\mu\varphi\partial^\mu\phi 
+ b_\mu b^\mu\varphi^2\Big] \nonumber\\&+& \frac{B^2}{2} + B\Bigl( \partial \cdot b + m^2 \varphi^2 \phi\Bigr)^2 \nonumber\\
&+& i\partial^\mu \bar c \partial_\mu c + i m^2 \varphi^2 \bar c c .\label{BRST22}
\eea
with scalar (anti-)ghost ($\bar c$)$c$ and Nakanishi-Lautrup auxiliary scalar $B$, with
$B=-\big( \partial \cdot b + m^2 \varphi^2 \phi\big)$ on-shell \cite{ASM}. The above Lagrangian, under
the (anti-)BRST transformations,
\bea
&& s^{(2)}_b h_{\mu\nu} = - 2c \eta_{\mu\nu}, \quad  s^{(2)}_b  b_\mu = \partial_\mu c, \quad  s^{(2)}_b \phi = \varphi c , \nonumber\\
&& s^{(2)}_b c = 0, \qquad \quad s^{(2)}_b \bar c = i B, \qquad   s^{(2)}_b B = 0, \nonumber\\
&& s^{(2)}_{ab} h_{\mu\nu} = - 2\bar c \eta_{\mu\nu}, \quad s^{(2)}_{ab} b_\mu = \partial_\mu \bar c, \qquad s^{(2)}_{ab} \phi = \varphi \bar c, \nonumber\\
&& s^{(2)}_{ab} \bar c = 0, \quad  s^{(2)}_{ab} c = - i B, \quad s^{(2)}_{ab} B = 0.\label{37}
\eea
changes by the total derivative terms,
\bea
s^{(2)}_b {\cal L}^{(2)}_{\rm GF2} &=& \partial_\mu \Bigl[ \frac{1}{24\kappa}  \bigl ( h \partial^\mu c -\partial^\mu h c - 4 c \partial_\nu h^{\mu\nu}\bigr)  \varphi^2 \nonumber\\
&+& \frac{1}{\kappa}\varphi \phi \partial_\mu c + B \partial^\mu c \Bigr],\nonumber\\
s^{(2)}_{ab} {\cal L}^{(2)}_{\rm GF2} &=& \partial_\mu \Bigl[ \frac{1}{24\kappa}  \bigl ( h \partial^\mu \bar c -\partial^\mu h \bar c 
- 4 \bar c \partial_\nu h^{\mu\nu}\bigr)  \varphi^2 \nonumber\\
&+& \frac{1}{\kappa}\varphi \phi \partial_\mu \bar c - B \partial^\mu \bar c \Bigr],\label{38}
\eea
respectively, leaving the corresponding action invariant.

As in the case of diffeomorphism before, only partial physical spectrum has been achieved,
as is evident from the corresponding equations of motion, with only Eq. (\ref{29}) being reproduced,
yielding positive energy states only for the vector field $b_\mu$. The fact that the gauge-fixing
condition in Eq. (\ref{30}) relates $b_\mu$ with the scalar field $\phi$ makes the 2 negative 
degrees of freedom of the (anti-)ghost scalars compensating {\it together} against the redundant 
combination of 4 positive degrees of freedom of $b_\mu$ and 1 of $\phi$. This leaves three
physical degrees of freedom for $b_\mu$, as $\phi$ does not represent a dynamical mode, analogous
to the St\"uckelberg theory \cite{Ruegg}, having a positive spectrum and thereby being the only
physical field, corresponding to the particular symmetry, coupled to $h_{\mu\nu}$ that contains
un-physical states presently.
Therefore, It is clear that though diffeomorphism and generalized Weyl gauge-fixings
are mutually independent, all the dynamical components (field) of the present theory cannot be
made physical, at the same time, by any one of them. A direct evidence of the same is the fact that, Eq.
(\ref{31}) for $\phi$ is {\it not} obtained in any of these `independent' procedure, and thus
the scalar degree of freedom remains ill-defined in both of them. This is because, in the second
order theory, $\phi$ is the only field, out of the three, that directly couples to the other two.
Therefore, the positive-norm Fock space for the full theory can only be achieved with both diffeomorphism and
generalized Weyl gauge-fixings applied, which is achieved below.

\subsection{For combined diffeomorphism and generalized Weyl symmetry}
As discussed above, in order to obtain the complete physical spectrum of the theory, both the gauge-fixing conditions
in Eqs. (\ref{28}) and (\ref{30}) are to be applied. Including the corresponding auxiliary 
Nakanishi-Lautrup and (anti-)ghost fields, of both vector and scalar types, the full tree-level quantum 
Lagrangian can be written as,
\bea
{\cal L}^{(2)}_{\rm GF}&=& \frac{1}{12\kappa}\Big[\frac{1}{4}h_{\mu\nu}\Box h^{\mu\nu}+\frac{1}{2}h \partial_\mu\partial_\nu h^{\mu\nu}
+\frac{1}{2} \partial_\alpha h^{\mu\alpha} \partial^\beta h_{\mu\beta} \nonumber\\
&-&\frac{1}{4}h\Box h\Big]\varphi^2
+\frac{1}{6\kappa}\Big[\partial_\mu\partial_\nu h^{\mu\nu} - \Box h\Big]\varphi\phi +\frac{1}{2\kappa}\partial_\mu \phi \partial^\mu \phi \nonumber\\
&-& \frac{1}{4}B_{\mu\nu}B^{\mu\nu}  +\frac{1}{2}m^2\Big[\partial_\mu\phi\partial^\mu\phi- 2 b_\mu\varphi\partial^\mu\phi 
+ b_\mu b^\mu\varphi^2\Big] \nonumber\\
&+& \frac{1}{24\kappa} B_\mu B^\mu + \frac{1}{12\kappa}B_\mu \Bigl(\partial_\nu h^{\mu\nu}-\frac{1}{2} \partial^\mu h
 - 2\varphi^{-1}\partial^\mu \phi \Bigr) \nonumber\\
&+& \frac{B^2}{2}  + B\Bigl( \partial \cdot b + m^2 \varphi^2 \phi\Bigr)^2 + \frac{i}{12\kappa} \partial^\mu {\bar c}_\nu \partial_\mu c^\nu \nonumber\\
&+& i\partial^\mu \bar c \partial_\mu c + i m^2 \varphi^2 \bar c c .\label{BRST}
\eea
Thus, the unphysical ghost sector, imbibing negative norm states, is completely decoupled from the
other dynamic fields, leaving them associated only with positive semi-definite energies, and
thereby removing the gauge redundancies due to {\it both} diffeomorphism and generalized Weyl
symmetries. The latter is now re-structured as the following BRST and anti-BRST transformations,
\bea
&&s_b h_{\mu\nu} = \partial_\mu c_\nu + \partial_\nu c_\mu - 2\eta_{\mu\nu} c, \quad s_b b_\mu = \partial_\mu c, \nonumber\\
&& s_b \phi = \varphi c,\quad s_b {\bar c}_\mu = i B_\mu, \quad s_b \bar c = i B, \nonumber\\ &&s_b [c_\mu,\; c,\; B_\mu,\; B] = 0,\nonumber\\
&&s_{ab} h_{\mu\nu} = \partial_\mu {\bar c}_\nu + \partial_\nu {\bar c}_\mu - 2\eta_{\mu\nu} c, \quad s_{ab} b_\mu = \partial_\mu c, \nonumber\\
&&s_{ab} \phi = \varphi c,\quad s_{ab}  c_\mu = - i B_\mu, \quad s_{ab} c = - i B,\nonumber\\
&&s_{ab} [{\bar c}_\mu,\; \bar c,\; B_\mu,\; B] = 0. 
\eea
Under these transformations, the Lagrangian ${\cal L}^{(2)}_{\rm GF}$ respectively transforms by
the total derivatives,
\bea
 s_b {\cal L}^{(2)}_{\rm GF} &=& \partial_\mu \Bigl[ \frac{1}{24\kappa} \Big\{(\partial_\nu c^\alpha) \partial^\mu h^{\nu\alpha} 
-  \partial^\mu (\partial_\nu c_\alpha) h^{\nu\alpha} \nonumber\\ 
&-& (\partial_\alpha c^\alpha) \partial^\mu h 
+  \partial^\mu(\partial_\alpha c^\alpha) h + 2 (\partial_\alpha c^\alpha) \partial_\nu h^{\mu\nu} \Big\}\varphi^2 \nonumber\\
&+& \frac{1}{\kappa}\varphi \phi \partial_\mu c
+  \frac{1}{24\kappa}  \bigl ( h \partial^\mu c -\partial^\mu h c - 4 c \partial_\nu h^{\mu\nu}\bigr)  \varphi^2\nonumber\\
 &+& \frac{1}{12\kappa}B^\nu \partial^\mu c_\nu + B \partial^\mu c \Bigr],\nonumber\\
s_{ab} {\cal L}^{(2)}_{\rm GF} &=& \partial_\mu\Bigl [\frac{1}{24\kappa} \Big\{(\partial_\nu {\bar c}^\alpha) \partial^\mu h^{\nu\alpha} 
-  \partial^\mu (\partial_\nu {\bar c}_\alpha) h^{\nu\alpha} \nonumber\\ &-& (\partial_\alpha {\bar c}^\alpha) \partial^\mu h 
+  \partial^\mu(\partial_\alpha {\bar c}^\alpha) h + 2 (\partial_\alpha {\bar c}^\alpha) \partial_\nu h^{\mu\nu} \Big\}\varphi^2 \nonumber\\
&+& \frac{1}{\kappa}\varphi \phi \partial_\mu \bar c  
 + \frac{1}{24\kappa}  \bigl ( h \partial^\mu \bar c -\partial^\mu h \bar c 
- 4 \bar c \partial_\nu h^{\mu\nu}\bigr)  \varphi^2 \nonumber\\
&-& \frac{1}{12\kappa}B^\nu \partial^\mu {\bar c}_\nu - B \partial^\mu \bar c \Bigr],\label{38}
\eea
leaving the action invariant. Both the relevant fields are now properly gauge-fixed, and satisfy `free' equations 
(\ref{27}), (\ref{29}) and (\ref{31}), which can lead to quantum effects, due to respective couplings.

Physically, it is crucial to evaluate the actual degrees of freedom of the complete
Lagrangian ${\cal L}^{(2)}_{\rm GF}$, in the (anti-)BRST formalism.
We count 10 tensor ($h_{\mu\nu}$), 4 vector ($b_\mu$) and 1 scalar ($\phi$) component of
dynamical fields, yielding 15 positive degrees of freedom (not considering the non-dynamic
auxiliary fields). The negative degree of freedom count is 10 (4 vector and 1 scalar each for
both ghost and anti-ghost kind). Therefore, the system has 5 physical degrees of freedom, 2 for 
$h_{\mu\nu}$ (graviton) and 3 for $b_\mu$ (massive gauge boson). This is the result we obtained 
in the last section, but at the level of equations of motion. This re-ensures the non-dynamical 
status of $\phi$, governed by Eq. (\ref{31}), as in the previous section.

Therefore, the physical domain of the STG-St\"ucke-lberg theory has been completely identified in
the weak field approximation. This is done through the (anti-) BRST procedures, ensuring the same
to be {\it off-shell}, thus enabling Euler-Lagrange equations derived out of the Lagrangian in Eq.
(\ref{BRST}) to yield proper tree-level propagators for both
$\bar{H}_{\mu\nu}$ and $b_\mu$. In the generalized graviton sector, the gauge-fixed EOM reads,
\bea
&&\frac{1}{2}\Big[\eta^{\mu\alpha}\eta^{\nu\beta}+\eta^{\mu\beta}\partial^\nu\partial^\alpha+\eta^{\nu\beta}\partial^\mu\partial^\alpha \nonumber\\
&&-\eta^{\mu\nu}\partial^\alpha\partial^\beta\Big]\bar{H}_{\alpha\beta}=0,
\eea  
which is exactly of the form for pure gravity under harmonic gauge \cite{Carroll}, yielding the 
corresponding propagator,
\be
\Delta^{\mu\nu,\alpha\beta}=-\frac{i}{2}\frac{\eta^{\mu\alpha}\eta^{\nu\beta}+\eta^{\mu\beta}\eta^{\nu\alpha}-\eta^{\mu\nu}\eta^{\alpha\beta}}{p^2+i\epsilon},\label{39}
\ee
representing an effective massless mode \cite{'tHV,Carroll}. Similarly, the gauge propagator is
similar to the St\"uckelberg propagator \cite{Ruegg},
\be
\Delta^{\mu\nu}=-i\frac{\eta^{\mu\nu}}{p^2-m_b^2}.\label{40}
\ee

The present gauge-fixing, of the quadratic Lagrangian, having both reduced
diffeomorphism and generalized Weyl symmetry, is of the most general form and complements the 
results in \cite{Faria,Pagani}, especially those with canonical approach \cite{Faria}. This is
sufficient to quantize the theory at the tree-level, with {\it bare} propagators specified. However,
the quantum corrections to a particular field, due to the interaction with another, can provide 
corrections to the bare structures, and even lead to anomalies. We leave such possibilities for 
future investigations.

\section{Conclusions}
In conclusion, the STG-St\"uckelberg theory with non-minimal scalar-gravity coupling,
following background field method, led to inconsistencies for a not-trivial
background metric; as {\it both} diffeomorphism, as well as generalized Weyl symmetry of the classical theory, are not respected. 
However, very interestingly, in Minkowskian background, this theory can be consistently gauge-fixed by retaining 
both the symmetries, where all the negative norm states (ghost states) are decoupled from the physical Fock space. Earlier, the quantization of
massive conformal gravity has been carried out by showing that the theory is renomrmalizable but has ghost states \cite{Faria}.
Here, the ghost-free Fock space has been achieved, after relating all symmetries; namely diffeomorphism and generalized
Weyl symmetry. Further, the graviton and massive gauge field represents the entire dynamics, and the
scalar field remain non-dynamical in the present case.

Using the well-established BRST technique, the qua-dratic Lagrangian, emerging from the
STG-St\"uckelberg theory, has been {\it independently} gauge-fixed, with 
respect to the diffeomorphism and generalized Weyl transformations.
In the gravitational sector, corresponding to diffeomorphism only, 
an appropriately {\it extended} harmonic gauge condition, of the de Donder form, is required, yielding a pair of vector (anti-)ghost fields. This leaves-out 
the negative norm states with respect to the dynamical vector field $b_\mu$ in the theory.
On the other hand, gauge-fixing only the extended Weyl gauge redundancy yields three physical degrees of freedom for the 
gauge field $b_\mu$, as the scalar field $\phi$ is not dynamical. Thus, negative norm states exist in both the cases, and the 
complete physical sector of the STG-St\"uckelberg theory is not identified, preventing the corresponding
quantization. The identification of the full physical Fock space of the quadratic STG-St\"uckelberg 
Lagrangian requires a vector and a scalar pair of (anti-)ghost terms, obtained by
{\it combining} the corresponding transformations. All the unphysical degrees of freedom of 
fields $h_{\mu\nu}$ and $b_\mu$  have been removed, yielding the 5 physical degrees of freedom, 2 for graviton ($h_{\mu\nu}$) and
3 for the massive gauge boson ($b_\mu$). It will be worthwhile to investigate effective theories
emerging from the present one, with interesting possibilities in areas like conformal gravity 
and gauge theories, as well as in condensed-matter systems. Some of these questions will be investigated in future works. \\

\end{document}